# Cascading Effects of the COVID-19 Pandemic on Barangays in the Philippines

Naomi Ashley Amparo, John Frederick Muji, Paul James Montecillo, Jaymar Soriano, and Vena Pearl Bongolan*
Systems Modelling and Simulation Laboratory
Department of Computer Science
University of the Philippines
1101 Diliman, Quezon City, Philippines

**Abstract**

The COVID-19 pandemic disrupted socio-economic and healthcare systems in the Philippines, significantly affecting barangays [9]. This study analyzes the cascading effects of the COVID-19 pandemic on key aspects of a barangay, namely mobility, accessibility of public services, economic and financial health, food security, educational engagement, and physical health. It focuses on data from 2,122 Filipino households collected during May to June 2021 as part of the World Bank COVID-19 Households Survey [15]. A Bayesian network model was constructed to programmatically map the conditional dependencies among these variables, utilizing Python libraries such as pyAgrum [11]. Survey responses were grouped into common variables based on shared characteristics and standardized through z-score normalization to serve as nodes in the Bayesian network. By extending the Bayesian network into an influence diagram, this study's findings will help identify interventions to guide local government units (LGUs) and policymakers in crafting tailored recovery programs and strategies that address impacts on physical health, economic and financial health, food security, public service access, mobility, and educational engagement. These efforts ultimately aim to enhance barangay resilience and preparedness for future public health crises. The results indicate that interventions aimed at boosting food production, stabilizing market prices, and expanding income opportunities are the most effective in improving community outcomes. This highlights the vital role of targeted economic and food security measures in mitigating the socio-economic impacts of the pandemic and offers valuable insights for shaping future response and recovery efforts.

**Keywords:** COVID-19 pandemic, barangays, bayesian network, local government units

# 1 Introduction

The COVID-19 pandemic has had a profound impact on the socio-economic and healthcare systems of the Philippines. It triggered widespread job losses, restricted access to essential public services, and led to deteriorating conditions in local communities, particularly at the barangay level. As a result, financial hardship, food insecurity, increased crime rates, and worsening physical and mental health became widespread challenges.

According to a study conducted by the World Bank [15], the pandemic caused substantial negative effects across multiple dimensions of Filipino households, including employment, income and livelihood, food security, health, education, and access to financial services. These impacts were marked by a tripling of the unemployment rate, income losses, and an economic contraction of up to 9%. Containment measures such as the implemented quarantines further restricted mobility, disrupted educational systems through prolonged school closures, and intensified the uncertainty regarding a return to face-to-face learning.

This study builds upon previous research titled "From the COVID-19 Pandemic to the Mental Health of the Philippines: Modelling the Cascade of Disasters using an Influence Diagram" [10]. While the earlier study focused on modeling the cascading effects of the pandemic on the mental health of Filipinos, this study aims to examine the broader ripple effects of the pandemic on key aspects of barangay life. Specifically, we investigate the impacts on mobility, accessibility of public services, economic and financial health, food security, educational engagement, and physical health

The study draws on data from 2,122 Filipino households collected between May and June 2021 as part of the World Bank COVID-19 Household Survey [15]. Our objective is to provide insights that will assist local government units (LGUs) and policymakers in formulating recovery programs and strategic interventions aimed at strengthening physical health, economic resilience, food security, public service delivery, mobility, and educational engagement in preparation for future public health crises.

Following the methodology of the previous study, we employ Bayesian networks to model the interrelationships among key variables affecting barangay outcomes. In addition, we extend the model into an influence diagram to identify optimal intervention strategies. Unlike the earlier work, this study goes further by using influence diagrams not just to map dependencies, but also to determine actionable interventions that can improve outcomes across multiple community dimensions.

# 2 Literature

### 2.1 COVID-19 and the Barangay

The COVID-19 pandemic severely impacted barangay communities, as they played a critical frontline role in implementing health protocols, distributing aid, and ensuring public order during times of crisis. During the pandemic, barangays became the center for local-level governance, where officials and community volunteers had to enforce lockdowns, manage relief operations, and provide critical information to residents [9]. A study by Dabalos et al. (2021) [7] surveyed barangay tanods in Bulacan and Metro Manila aged 38 to 55 years old to determine how they adapted to the changes in their work circumstances brought about by the pandemic, and how these experiences could improve policies in future crises. The study's results showed that the barangay tanods had a hard time adjusting to the sudden shift in their work functions. For instance, the pandemic added to their workload: instead of simply doing day and night patrols, they now had to guard checkpoints and monitor compliance with health protocols. They also expressed apprehension about potentially being afflicted with COVID-19 every time they stepped outside to fulfill their duties. Despite the exhaustion from the additional workload, they still managed to remain optimistic and did not fail to perform their duties effectively. Most of the tanods also received incentives from the government to support their health and sanitary needs.

A study by Cabatac and Java (2024) [3] determined the difficulties faced by members of various sectors such as health, agriculture, business, private, and government in rural parts of the country. The study also examined the strategies they employed to cope with these difficulties and the lessons they learned. The challenges included travel restrictions, limited mobility, limited social interactions, and the prohibition of celebrating special occasions. Specifically, the agricultural sector found the enhanced community quarantine very difficult, as only one person from the family was allowed to go outside to buy necessities for the farm. The business sector faced difficulties in travel due to restrictions, which hindered their need for daily routines such as processing documents. The education sector struggled to maintain social distancing while at work. Meanwhile, local residents found it difficult to celebrate special occasions due to social distancing measures.

The health sector expressed significant difficulty in living with immunocompromised family members, such as senior citizens and infants, fearing they could transmit the virus to them. To cope, sectors employed various mechanisms, such as engaging in personal hobbies, praying, following strict personal protective measures, and avoiding public places. Specifically, the agricultural, health, and education sectors practiced strict measures like wearing masks and regular hand washing when outside and at work. The business sector focused on motivating themselves to face the outbreak optimistically. Local residents often turned to prayer, worship, and Bible study as coping strategies.

In summary, the participants across different sectors shared lessons learned from the crisis, including the importance of creating budgetary plans to accommodate more limited resources. They also emphasized trust in the government's protocols, believing that everyone must strictly adhere to health guidelines to overcome the pandemic. Furthermore, the participants recognized the necessity of vaccination to return to the new normal. The participants' faith in Christianity served as a crucial pillar in their survival during the crisis, highlighting prayer as a vital source of strength.

Taking into account the hardships faced by all the sectors, it further emphasizes that Razu et.al., (2021) [12] stated that healthcare services in rural areas faced tremendous difficulty dealing with the situation, experiencing uncertainty, anxiety, and fear. In addition, individuals encountered challenges such as stigmatization, heavy workloads, shortages of quality personal protective equipment, lack of incentives, poor coordination, and psychological stress .

The World Bank in collaboration with the Department of Social Welfare and Development (DSWD) conducted a High Frequency Social Monitoring survey on the 101 barangays in the country, particularly most vulnerable communities, to better understand the socio-economic impacts of pandemic on a community level. They found the most pressing problems for the community before and after the pandemic were the lack of income opportunities and reduction of pay, insufficient food supply, and health, sanitation and nutrition. The communities experienced severe job losses in more than 50% of the respondent communities, primarily in the sectors of construction and public transportation. Issues in food supply nearly doubled from 17% to 40% during the pandemic, while health and sanitation aspects remained consistent before and after COVID-19, but with the onset of the pandemic, other challenges were reported including lack of personal protective equipment and medical supplies and a lack of health facilities for COVID-19 cases. After the pandemic, poverty also remained a persistent problem for the communities. The survey also showed that 23% of the farmers, as well as the self-employed or family-owned businesses, reported declining revenues. At the barangay level, 32% of the respondent communities found that barangay services have been negatively affected, particularly that of health services and social assistance. IP communities are reported to have been affected by these effects more severely than the overall sample of communities given that they were remote and have constrained access to the nearest large city. Overall, the survey demonstrates that poor communities experienced significant economic impact from the pandemic [15].

### 2.2 Mitigations

The national government implemented several blanket policies to control the spread of COVID-19, including voluntary physical distancing, mandatory face masks and face shields, mass testing, and school closures. However, the effectiveness of these measures varied across communities. In contrast, local government unit (LGU) interventions were more granular, allowing them to better address the heterogeneity of local contexts.

During the pandemic, LGUs adopted a range of non-pharmaceutical containment measures to mitigate viral transmission. These included restricting movement by age group and allowing only essential businesses to operate. According to a study conducted by Talabis et al. (2021) [14], effective LGU-level responses included strict border control, early lockdown implementation, establishment of quarantine facilities, and clear public communication strategies.

Villanueva and Buhat (2020) [2] developed an agent-based model to simulate infectious disease spread within a small community. By adjusting parameters such as population size, infection and recovery rates, and asymptomatic cases, they identified key interventions that significantly reduced mortality and infection rates. Random mass testing, social distancing, and targeted quarantines emerged as effective strategies. Their simulations found that the most feasible intervention was the implementation of enhanced community quarantine combined with random testing of up to 100 individuals.

Agdeppa et al. (2022) [1] examined how community quarantines led to food insecurity among Filipino families. Contributing factors included loss of income, lack of public transportation, mobility restrictions, and absence of younger household members capable of procuring food. Families adopted various coping strategies such as buying food on credit, borrowing, bartering, reducing adult food intake, pawning assets, and seeking help from LGUs or formal institutions. Intervention programs that families accessed included food and cash assistance programs.

To accelerate digital inclusion during the pandemic, community-level initiatives were undertaken as part of the National Broadband Plan. These included infrastructure development to provide fast, reliable, and affordable broadband services, deployment of solar-powered electricity, repair and maintenance of ICT equipment, and training programs in digital literacy and cybersecurity [13].

Cordero (2023) [5] highlighted the role of telemedicine as an effective alternative to in-person consultations. Telemedicine enabled remote access to medical professionals via smartphones, hotlines, websites, and mobile applications. Patients could undergo consultations, receive diagnoses, and obtain prescriptions virtually. In the Philippines, this was supported by legislative efforts such as the Philippine E-Health and Telemedicine Development Act of 2020, filed by Representative Joey Salceda. The use of telemedicine was associated with reduced infection risk, improved chronic disease management, and enhanced family engagement in care.

To support continued education, the Department of Education (DepEd) issued guidelines for remedial and advancement classes in the summer of 2021. These were delivered through distance learning modalities tailored to each community's health context. Instruction was provided using Self-Learning Modules (SLMs) or Alternative Delivery Modes (ADM), focusing on helping learners meet the Most Essential Learning Competencies (MELCs) for their grade levels [8].

# 3 Methodology

### 3.1 Influence Diagram

#### 3.1.1 Chance Nodes

A Bayesian network model was used to analyze the cascading effects of COVID-19 on the state of a barangay. The model maps the conditional dependencies between variables representing the key aspects of local communities affected by the pandemic. This process was programmatically using python libraries, including pyAgrum [11]. The Bayesian network was formulated using the data from Round 3 of the World Bank COVID-19 Households Survey [15]. Survey questions were grouped by shared characteristics and standardized by their z scores to be used as nodes for the Bayesian network. Notably, not all survey questions were utilized, but rather only those that were deemed relevant to the study were used.

**First Layer**

- Government Response - The nodes grouped in this category are based from the survey questions measuring how satisfied the community members are in terms of the COVID-19 response of the government on a local and national level. It also includes the participants' ratings on the efficacy of the government's response as well as its trustworthiness as an institution.
- Government Assistance - These nodes in the survey indicate whether the participants received government assistance, and if so, how much assistance they received.

**Second Layer**

- COVID-19 Information and Vaccine - This group of nodes covers the availability of COVID-19 vaccines and COVID-19 information. Whether there is enough vaccine supply, if participants have concerns about getting vaccinated, whether they have been vaccinated, and whether they have been tested for COVID-19.
- Mobility - This covers the question of whether the participant experienced more restricted mobility in doing their usual activities.

**Third Layer**

- Economic Impact - This group of nodes cover the participants' employment status, their personal reasons as to why they are not working, the effect of the pandemic on their income, whether they still receive remittances from OFW relatives, and their degree of financial worry.
- Health - This covers whether the participants have needed and/or accessed medical treatment during the pandemic.

**Fourth Layer**

- Accessibility - This group of nodes assess the number of smartphones each household has and whether they have access to the internet.
- Food Security - These nodes refer to questions inquiring into the reasons participants could not buy staple foods like rice, meat, fruits, and vegetables
- Educational Impact - This covers the number of children enrolled in each household and their interest in resuming face-to-face learning.

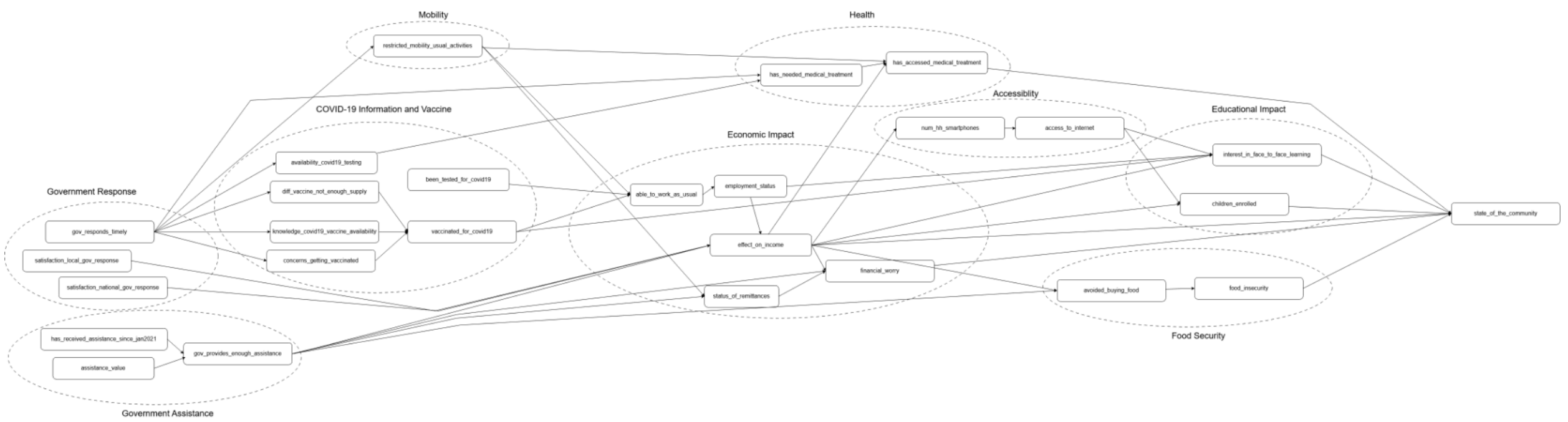

Figure 1: Bayesian influence diagram of cascading effects of COVID-19 on the state of a barangay

### 3.1.2 Decision Nodes

The decision nodes in this study were derived from consultations with subject-matter experts who had direct experience with the interventions, mitigations, and policies implemented during the COVID-19 pandemic. These experts included a labor economist, a local government policy officer, personnel from the Department of the Interior and Local Government (DILG), a barangay health officer, a public school teacher, and staff from the Department of Agriculture. Collectively, they provided insight into key aspects of barangay life, including mobility, access to public services, economic and financial well-being, food security, educational engagement, and physical health.

Table 1 presents the interventions for the decision nodes, their corresponding descriptions, and the chance nodes they influence. The efficacies of most interventions were estimated by the consultants using their domain knowledge and experiential judgment regarding how each intervention affected outcomes in their respective areas. Notably, the decision nodes, Additional Online Classes, Multi Modal Distance Learning, Agriculture Livelihood Support, Enhance Food Production, Video Lessons, Market Price Stabilization, and Food Distribution Accessibility, were estimated by the researcher to have an efficacy of 100% as the consultant was unable to provide their estimated efficacies.

Table 1: Efficacies, descriptions, and impacts of decision nodes

| Decision Node | Efficacy | Description | Impacts |
|---|---|---|---|
| Expanded Vaccination Program | 95% | Increasing the reach of vaccination programs by engaging with private entities and local leaders. | Positive impact on variable 'vaccinated for COVID-19' |
| Isolation Facilities | 100% | Provision of local isolation facilities to prevent further infections in the barangay. | Negative impact on 'has needed medical treatment'. |
| Frontline Support | 100% | Support for vulnerable groups(elderly, persons with disabilities, and low-income families) through supplies and health monitoring. | Positive impact on 'has accessed medical treatment'. Negative impact on 'food insecurity'. |
| Localized Lockdown Coordination | 85% | Targeted containment with regular reports and clear communication to ensure adherence and effective lockdown. | Positive impact on 'government response is timely'. |
| Health Education | 80% | Education of the community about COVID-19 prevention measures and vaccination and dispelled vaccine myths and fears. | Positive impact on 'knowledge about COVID-19 accessibility' and negative impact on 'concerns about getting vaccinated'. |
| Community Mobilization | 80% | Motivation of community participation in health programs by providing knowledge and improving the health and well-being of community members. | Positive impact on 'has accessed medical treatment' and 'vaccinated for COVID-19'. |
| Hotline System | 80% | Reliable platform to seek information, reducing misinformation and confusion. | Positive impact on ‘government response is timely’ and ‘satisfaction to local government response’. |
| Mobile Market | 85% | Market moving from barangay to barangay to improve access to vegetables and other food items. | Negative impact on ‘food insecurity’. |
| Additional Online Classes | 100% | Online classes for those with gadgets and internet connection. | Positive impact on ‘children enrolled’ and ‘access to internet’. |
| Multi Modal Distance Learning | 100% | Learners receive printed modules barangay officials/stakeholders deliver. Weekly outputs are retrieved. Continues education system with minimal face-to-face contact. | Positive impact on ‘children enrolled’ and ‘interest in face to face learning’. |
| Agriculture Livelihood Support | 100% | Help marginalized farmers regain the capacity to continue agricultural activities through capital and financial stability | Positive impact on ‘effect on income’. |
| Enhanced Food Production | 100% | Programs to ensure food supply availability, accessibility and price stability. Increase local production for households to rely less on external supply chains. Reduces risk of food shortages and safeguards the local economy. | Positive impact on ‘effect on income’ and negative impact on ‘food insecurity’. |
| Video Lessons | 100% | Pre-recorded video lessons or DepEd TV segments for each subject/grade level. Allows for wider reach, only needing TV/cable or USB copies. | Positive impact on ‘children enrolled’ and ‘interest in face to face learning’. |
| Market Price Stabilization | 100% | Prevent profiteering and hoarding to aid low-income households in accessing food. | Negative impact on ‘food insecurity’ and ‘financial worry’. |
| Food Distribution Accessibility | 100% | Help farmers deliver goods to markets and consumers. Reduces supply chain disruptions. | Negative impact on ‘food insecurity’. |
| Cash for Work | 90% | Provides immediate income for displaced workers to boost economic and psychosocial wellbeing. | Positive impact on ‘effect on income’ and negative impact on ‘financial worry’. |
| Livelihood Reskilling | 80% | Empowers unemployed or displaced workers, especially women, with new income streams (e.g., food processing, sewing, community gardening). | Positive impact on ‘employment status’ and ‘effect on income’. |
| Frontline Educators Protections | 75% | Ensures teachers and para-teachers, especially in public and Lumad schools, have sufficient protection to resume or continue duties. | Positive impact on ‘able to work as usual’ and ‘interest in face to face learning’. |
| Barangay Labor Rights Awareness | 75% | Establishes safe local spaces for reporting labor violations, including illegal dismissals, wage theft, and lack of benefits. | Positive impact on ‘able to work as usual’ and ‘employment status’. |
| Expanded BHW Livelihood Support | 70% | Strengthens health systems by formally recognizing and compensating BHWs. | Positive impact on ‘able to work as usual’ and ‘has accessed medical treatment’. |
| Community-based Daycare | 80% | Frees up unpaid caregivers (often women) to rejoin the workforce and improve elder welfare. | Positive impact on ‘employment status’. |
| Reintegration of OFW | 75% | Supports returning migrant workers through livelihood grants, skills matching, and psychosocial services. | Positive impact on ‘able to work as usual’ and ‘status of remittances’. |
| Mandatory Masks | 100% | Mandatory wearing of face masks to reduce transmission of COVID-19. | Positive impact on ‘able to work as usual’ and negative impact on ‘has needed medical treatment’. |
| Health Worker Shelters | 100% | Temporary accommodations for the safety and protection against discrimination of health workers. | Positive impact on ‘has accessed medical treatment’, and ‘satisfaction of local government response’. |

| Food Security Nutrition Program | 100% | Programs to promote public awareness, participation, and coordinated action to address hunger, improve nutrition, and ensure food security. | Negative impact on ‘food insecurity’. |
|---|---|---|---|

**3.1.3 Utility Node**

The utility node is the state of the community, chosen because the study’s goal is to identify the optimal decisions for mitigation plans in issues arising from the key aspects of barangay life-educational engagement, economic impact, health services, food security, and mobility. The node is encoded with three ordered states of low, medium, and high. After evidence is propagated through the network, each state receives a posterior probability that considers the rest of the network and subsequent calculations. These probabilities are then translated by pyAgrum into a single expected-utility score that helps measure how interventions impact the network and identifies the most beneficial decision nodes.

**3.2. Optimal Decisions**

To use the model as concrete policy guidance, we modeled the network in python (version 3.10) and performed all numerical inference using the pyAgrum library. The 2,122 household records from Round 3 of the World Bank COVID‑19 Household Survey were first re‑scaled to two‑state (low, high) or three‑state (low, medium, high) ordinal scale, depending on the distribution and conceptual granularity of the original question. This approach allows as much information to be preserved for when variables are naturally binary or when finer details are available. The preprocessed dataset was then used to inform each conditional‑probability table (CPT) that establishes the Bayesian network. The structure is turned into an influence diagram by appending the decision and utility nodes described in Sections 3.1.2 and 3.1.3.

Each decision node was encoded with the binary states yes (the intervention is implemented) and no (existing conditions remain). When a node is set to no, the child chance nodes preserve the learned CPTs. When an intervention is set to yes, the CPTs of its child chance nodes are perturbed by an amount equal to the intervention’s efficacy **e**. For a positive relationship (e.g., “Mobile Market” improving food access), a fraction equal to the efficacy **e** is subtracted proportionally from the low and medium probability masses and added to the high state (Only low state for binary state chance nodes). On the contrary, for a negative relationship (e.g., “Enhanced Food Production” decreasing food insecurity), the same fraction is shifted towards the low state.

For each decision node, the algorithm computes the posterior distribution of the “State of the Community” utility node and returns its expected utility, the probability that the community attains a high state. The policy (or policies) that maximizes this quantity represents the maximum expected utility (MEU) solution and therefore constitutes the model’s optimal decisions.

# 4 Results

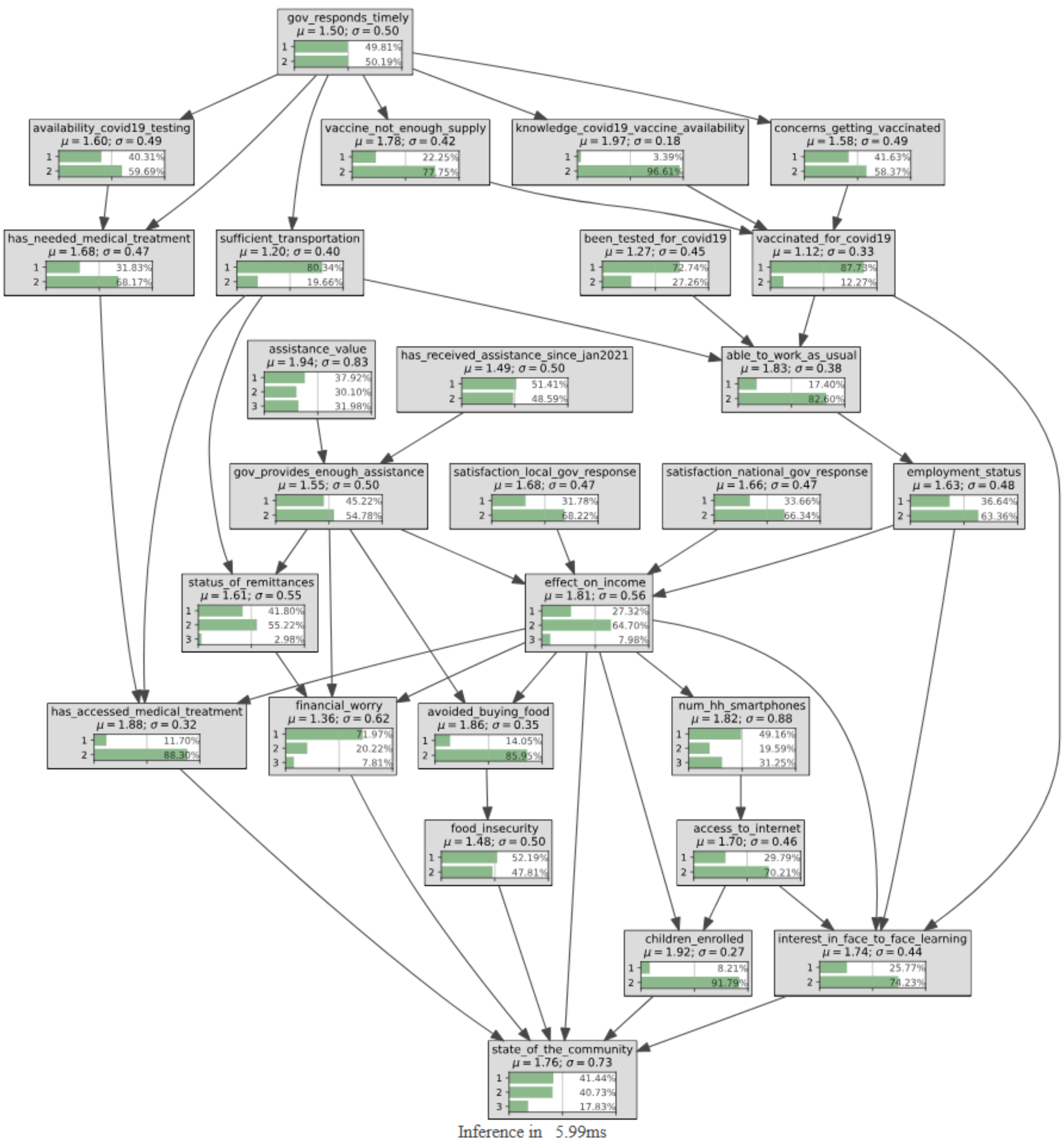

Figure 2: Bayesian influence diagram conditional probabilities

Figure 2 shows a visualization of the prior conditional probability tables learned from the June 2021 World Bank household survey. The network was trained in pyAgrum after raw questionnaire items were standardized then categorized to two- or three-state variables. During training, each node's probability distribution is iteratively adjusted and re-weighted to remain internally consistent with the joint probabilities of its parents and children. The result is a conditional-probability table in each node after the algorithm has reconciled it with every other variable and state in the network. This allows for the most accurate prior estimates that constitute the baseline state of the community from which our evidence and interventions can be applied.

One pattern that emerges is the dominance of the economic variables in the middle layer: 'effect on income' showcasing a combined 92.02% on its two worst states, cascading into financial worry with 92.19% and food insecurity with its worst state at 52.19%. The path to food insecurity is further moderated by the chance node of 'Avoided buying food' (14.05% yes), which indicates the effectiveness of coping mechanisms in buffering the link between falling income and hunger. Because 'effect on income', 'financial worry', and 'food insecurity' are currently concentrated in their least-desirable states, any shift toward better outcomes along these paths have substantial room to lift the overall community score. Moreover, with all five direct parents of the utility node drawing at least one arrow from the economic variables, interventions that affect income are expected to see an undiminished utility increase. In contrast, education nodes are unexpectedly healthy with the 'children enrolled' variable at 91.79% and 'interest in face to face learning' at 74.23% in their positive states. The CPT consequently assigns these variables a smaller weight; even policies with high efficacy here are expected to produce only modest changes in utility.

A further observation that completes the baseline network established by the CPTs is the presence of the government-support, health-services, and COVID-prevention sub-networks that occupy long, low-leverage paths. Both 'vaccinated for COVID-19' and 'been tested for COVID-19' are skewed toward their poorer states, yet their principal children- 'able to work as usual (worst state at 17.40 %)' and 'interest in face-to-face learning' (worst state at 25.77%)- are already near-optimal and according to the learned CPTs, have modest weight on utility. The case is similar in the health services and government support nodes: the heavily positive 'needed medical treatment' (88.3% "no") flows into the equally favourable 'accessed medical treatment', while the various government assistance-related nodes lose much of their signal before reaching high impact variables like financial worry or effect on income. Additional programmes that are within the health and government sectors(for example, isolation facilities or hotlines) can therefore move probabilities only marginally, a point that will resurface in the analysis of optimal decisions.

With the posterior probabilities of the Bayesian Network set, the interventions are entered individually as decision nodes to observe the changes in posteriors as recomputed by pyAgrum. Each intervention fixes its efficacy on one or more chance node states, and the package recalculates the joint distribution so every parent and child in the network remains consistent with new evidence. The intervention is recognized as welfare-improving when its input shifts probability mass from the worst states (1-2) to the best state (3) in the utility node, 'state of the community.'

Table 2: Optimal decisions and MEUs based on decision scenarios.

| Decision Node | Optimal Decision | MEU if decision is 'no' | MEU if decision is 'yes' |
|---|---|---|---|
| Market Price Stabilization | yes | 0.1783 | 0.8203 |
| Enhanced Food Production | yes | 0.1783 | 0.8105 |
| Cash for Work | yes | 0.1783 | 0.5773 |
| Agriculture Livelihood Support | yes | 0.1783 | 0.5709 |
| Livelihood Reskilling | yes | 0.1783 | 0.4711 |
| Frontline Support | yes | 0.1783 | 0.2909 |
| Food Security Nutrition Program | yes | 0.1783 | 0.2813 |
| Food Distribution Accessibility | yes | 0.1783 | 0.2813 |
| Mobile Market | yes | 0.1783 | 0.2658 |
| Reintegration of OFW | yes | 0.1783 | 0.2570 |
| Health Worker Shelters | yes | 0.1783 | 0.1839 |
| Multi Modal Distance Learning | yes | 0.1783 | 0.1826 |
| Video Lessons | yes | 0.1783 | 0.1826 |
| Frontline Educators Protections | yes | 0.1783 | 0.1824 |
| Community-based Daycare | yes | 0.1783 | 0.1822 |
| Barangay Labor Rights Awareness | yes | 0.1783 | 0.1820 |
| Mandatory Masks | yes | 0.1783 | 0.1817 |
| Isolation Facilities | yes | 0.1783 | 0.1813 |
| Community Mobilization | yes | 0.1783 | 0.1812 |
| Expanded BHW Livelihood Support | yes | 0.1783 | 0.1810 |
| Localized Lockdown Coordination | yes | 0.1783 | 0.1801 |
| Hotline System | yes | 0.1783 | 0.1801 |
| Expanded Vaccination Program | yes | 0.1783 | 0.1785 |
| Health Education | yes | 0.1783 | 0.1783 |
| Additional Online Classes | no | 0.1783 | 0.1774 |

The influence-diagram results upon inputting the decision nodes produced a clear hierarchy. Programs that inject resources at the very start of the economic-and-food sequence such as enhanced food production, price stabilization, and cash-for-work pushed our community-state score from about 0.18 to well above 0.57, with a couple of food-supply measures inching past 0.80. These policies are wired high in the graph, feeding directly into income and food-security nodes that, in turn, feed into the nodes: remittances, financial worry, children's schooling and finally the composite "state of the community."

Because each step multiplies rather than merely adds benefits, the initial boost survives the long trip to the utility node.

Decision nodes including frontline food distribution, mobile markets and support for returning migrant workers have gains that hover around 0.26 to 0.29. They still work toward providing food and money to the people, but they enter the network one hop later, so part of the signal is already lost when it reaches the deeper layers. This implies the positive effects have already dissipated by the time it reaches the utility node.

All remaining interventions add less than 0.02 to utility. Even highly efficacious measures such as Expanded Vaccination or Mandatory Masks appear weak because they are connected deep in the graph. Their influence passes through several noisy and low impact nodes that dilute the original positive impact.

At the opposite extreme, interventions related to education see a negligible gain in utility despite being close to the utility node. Although 'children enrolled' and 'interest in face-to-face learning' are direct parents of the utility node, the CPT learned from the survey gives them small coefficients. Consequently, interventions with 100% efficacy like 'Video Lessons' only adds +0.043 in MEU while 'Additional Online Classes' lowers it by 0.0009. In this case, neither structural placement or efficacy became the bottleneck, and it was instead the learned CPT weights that decided the pay-off. These results from the interventions and CPT suggests that educational engagement was already high in June 2021 in the Philippines, and that further interventions only see marginal improvements.

# 5 Conclusion

The COVID-19 pandemic had cascading effects on the socio-economic aspect of barangays. This study modeled these impacts using Bayesian networks, which were extended into influence diagrams. Moreover, this study can be considered novel in that it does not just describe correlations, but it also models multi-sector community impacts and identifies optimal interventions via influence diagrams.

From the findings, economic and food-security-related interventions produced the largest improvements in the overall state of the community. For example, Enhanced Food Production, Market Price Stabilization, and Cash-for-Work led to significant utility gains ranging from 0.33 to 0.64. Additionally, Frontline distribution and livelihood support interventions had moderate effects with utility gains ranging from 0.26 to 0.29. On the other hand, while education and health interventions are deemed to be important, they had limited impact in the model used due to already favorable baseline conditions or lower CPT weights.

Furthermore, some high-efficacy interventions e.g. Expanded Vaccination only showed marginal utility gains, as their position in the network caused signal dilution through low-impact or low-variance nodes. Thus, the placement of an intervention in the network and the learned weight of its downstream effects also influence the outcome, showing that efficacy alone is not sufficient.

The study demonstrates the practical application of Bayesian influence diagrams in policymaking, especially for community-level resilience planning. The study can serve as a framework for data-driven prioritization of limited resources. Additionally, LGUs and national policymakers can use similar models to test intervention scenarios before implementation, to forecast policy outcomes when under uncertainty, and to design localized, high-leverage recovery plans.

One limitation of the study is that some CPT weights may be influenced by timing e.g., education already being high in June 2021. Additionally, expert-estimated efficacies are subjective in nature. In future work, we recommend expanding the model to include interaction effects between decision nodes as well as replicating the model in other countries or communities. We also recommend getting more data to increase the study's generalizability.

In conclusion, the model has potential as a policy-support tool that can integrate community-sourced data, expert insight, and probabilistic reasoning. As communities in the Philippines prepare for future health emergencies, tools like these offer a means to design smarter, faster, and more equitable interventions.

**Appendix A. Consultant Letters and Expert Essays**

The following letters and expert submissions were provided by consultants familiar with barangay-level pandemic interventions and were used to support the modeling of decision nodes in this study.

**Appendix A.1. Mitigations from Consultant 1**

Consultant Name: Father Mario Dimapilis
Affiliation:
Department of Nursing
St. Anthony's College
San Jose, Antique
Consultant Expertise: Barangay Health Worker
Title: Heroes in the Trenches: How Barangay Health Workers Battled the COVID-19 Pandemic in the Philippines

| Mitigation Measures/Decision Nodes | Efficacy | Impacts |
|---|---|---|
| 1. Building and maintenance of the community's temporary isolation facilities and quarantine areas in the barangay<br><br>(https://blogs.worldbank.org/en/eastasiapacific/strength-community-tackling-covid-19-recovery-philippines-through-community-driven) | 100% | Not every barangay in the Philippines had buildings, establishments, and ready accommodations for those who needed to be isolated to prevent the spread of the COVID-19 virus. The Barangay Health Workers helped to find a location far enough from where the houses were but readily accessible by ambulance and other vehicles to build the temporary isolation area in the barangay. They helped set up isolation facilities from local materials available in their communities. Since they knew how to contain the virus, they saw that the provisions and amenities in the temporary isolation areas prevented the spread of infection and cross-infection. They had to maintain these facilities throughout the long duration of COVID-19 emergencies. They also acted as primary caregivers of the people under isolation and quarantine. |
| 2. Frontline support for Vulnerable Neighbors<br><br>https://www.upm.edu.ph/cpt_news/bhws-share-stories-as-covid-19-frontliners/ | 100% | Community Health Volunteers and barangay health workers supported vulnerable groups, such as the elderly, persons with disabilities, and low-income families, by delivering essential supplies like food, medicine, and hygiene kits. They were the health front liners in their respective barangays. They conducted health teachings for their neighbors who were |
| | | under monitoring and health promotion campaigns for other community members in a door-to-door manner. They also played a crucial role in contact tracing by identifying and monitoring individuals who had been in contact with COVID-19-positive cases. |
| 3. Health Education<br><br>https://journals.plos.org/globalpublichealth/article?id=10.1371/journal.pgph.0000742 | 80% | They educated the community about COVID-19 prevention measures and vaccination and dispelled vaccine myths and fears. Despite their hard work in providing education, they also encountered resistance, which led to a vaccine hesitancy rate (56.6%). |
| 4. Community Mobilization<br><br>https://bmchealthservres.biomedcentral.com/articles/10.1186/s12913-020-05699-0 | 80% | Community Health Workers motivated and sustained community participation in health programs by providing knowledge and improving the health and well-being of community members |

**Appendix A.2. Mitigations from Consultant 2**

Consultant Name: Ron Herbon, Dr. Andrew Villacorta (from National Rice Program), and Ms. Perla Gines
Affiliation: DA DRRMS (Disaster Risk Reduction and Management Section)
Consultant Expertise: Agriculture and food security
Title: DA Interventions during COVID-19 Pandemic - for UP Research Team

During the COVID-19 pandemic, food has been available. However, one challenge in food security is accessibility as a result of restrictions imposed by the National and Local Government Units that hampered the passage of food products in the country.

The Department of Agriculture, together with partners from other agencies and the private sector, performed vital services and focused on key programs to help ensure food supply availability, accessibility and food price stability throughout the country during the pandemic.

**A. FOOD PRODUCTIVITY AND AVAILABILITY**

- **Rice Resiliency Project**
  This involves distribution of high quality seeds and fertilizers to farmers.

- **Urban Agriculture for Crops and Livestock**
  This program aims to (1) address food shortage in the community; (2) ensure adequate supply of readily available safe and nutritious food within the community; (3) Improve the environmental status of the community; and stabilizing food prices and strengthening market potential. With this strategy, food production will no longer be limited to rural areas only. Hence, urban dwellers may now grow their own food right in their backyards or in any available spaces around their households.

- **Direct Procurement from Farmers by LGUs**
  This involves Local Government Units (LGUs) buying agricultural produce directly to farmers for distribution to their constituents

**B. FOOD ACCESSIBILITY AND MOBILITY**

- **KADIWA ni Ani at Kita**
  A marketing program of the DA where farmer produce are directly sold to consumers. It eliminates the middlemen, making the farmers get the best prices for their goods, while providing adequate access, affordable, safe and nutritious food to the consumers.

- **Issuance of Food Passes**
  In line with the directives of the Interagency Task Force for the Management of Emerging Infectious Diseases (IATF-EID), the Department issued IATF IDs, Rapid Passes and Food Passes to ensure unhampered movement of all cargoes, agriculture and fishery inputs, food products, and agribusiness personnel nationwide amidst the quarantine.

**C. Food Price Stability**

- **Enforced Price Freeze for Basic Commodities**
  This was implemented through a Joint Memorandum Circular with DA, DTI and DOH on "Price Freeze under a State of Calamity throughout the Philippines throughout the Philippines due to Corona Virus Disease 2019 (COVID-19)" to ensure stabliity of prices for basic commodities amidst the pandemic

- **Partnered with the Local Price Coordinating Councils (LPSCCs) to strengthen price monitoring**
  This involves conduct of price monitoring of agricultural and fishery commodities in public markets by the Department of Agriculture Central Office and its Regional Field Offices, in partnership with the Local Price Coordinating Councils (LPCCs)

**D. SUSTAINABLE SOCIAL AMELIORATION**

- **Expanded SURE Aid and Recovery Program for Marginalized Small Farmers and Fishers (MSFF) and Micro and Small Enterprises (MSEs)**
  The project aims to: 1) finance the emergency and production capital requirements of marginalized, small farmers and fisherfolk (MSFF) whose incomes were affected by the enhanced community quarantine (ECQ) due to COVID-19 to help them regain capacity to continue their agricultural activities and contribute to sustained food production; and 2) provide the working capital requirement of micro and small enterprises (MSEs) engaged in agriculture and fisheries food production and delivery of produce/products/commodities to ensure the availability of food supply.

- **Rice Farmers Financial Assistance**
  This program aims to cushion the effects of sudden drop of palay price through giving an unconditional cash transfer in the amount of PhP 5,000.00 to eligible RSBSA-registered rice farmers tilling 0.5 to 2.0 hectares.

- **Financial Subsidy for Rice Farmers**
  Meanwhile, this program aims to cushion the effects of the sudden drop of palay price through giving an unconditional cash transfer in the amount of PhP 5,000.00 to eligible RSBSA-registered rice farmers tilling 1 hectare and below.

- Such programs were implemented in pursuant to Republic Act (RA) No. 11469 or the "Bayanihan to Heal as One Act"

### Appendix A.3. Mitigations from Consultant 3

Consultant Name: Kristin Mae S. Aala
Consultant Expertise: Public school teacher at Tagaytay City Science National High School-Integrated Senior High School
Title: None

COVID-19 had a huge effect in the education system of our country. It tested our commitment, our resilience and our love for education. Schools were forced to close indefinitely to prevent the spread of the virus. School closure brings difficulties for students, teachers, and parents. We made a lot of adjustments in order to adapt in the situation. The Department of Education did not stop planning and formulating different contingency plans that will address the problems occurred in the pandemic era. They were able to reconstruct and redesign the curriculum; they also designed various strategies to recover the gap and the lost learning, and provide learners and parents the assurance that even if face-to-face classes were suspended temporarily, the learning process will continue.

On the first phase of the pandemic, our school implemented the modular distance learning. In this time, we tapped the help of our stakeholders. The parents, private and local institutions who are willing and capable to help, and of course, the barangay councils served as our education heroes. We were able to deliver the modules of the learners together with the school supplies they need with the help of the barangay officials in their respective home addresses. The teachers were assigned to create a Weekly Home Learning Plan that will serve as a guide for the learners' tasks to do. The learners were given enough time to accomplish their assigned learning tasks. After a week, our education heroes will retrieve the outputs of the learners and hand it back to our school for assessing. Indeed, COVID-19 did not stop the learning process. Unfortunately, some learners and parents gave negative feedbacks on this alternative learning strategy that we developed. Distance learning is a solution to continue the education system, but it is difficult in our countries because many parents have not themselves been to school and there is a lack of the necessary Information and Communication Technology (ICT) infrastructures, computers, radio, and television to provide distance learning. There are learners who cannot accomplish their learning tasks alone at home, thus, they tend to just not do their assigned activities. And with this, our school needs to design other contingency plan in order to sustain the learning process.

On the second phase of the pandemic, we added another distance learning strategy. We offered online classes for learners who are capable of providing gadgets and strong internet connection. In order for the teachers to be digitally prepared, we undergo different online seminars in order for us to adjust and be familiar with the different online platforms that we may use in conducting our classes. But, negative feedbacks also arise in this new mode of distance learning. Teachers struggle with difficulties in the area of technology and lack of infrastructure availability.
Another option that we had during this pandemic is creating video lessons. In this option, teachers were tasked to create comprehensive video lessons that will teach the learners, give key concepts of the lesson, and administer different activities related to the lesson. The Department of Education also sponsored quality video lessons congruent to the curriculum aired in the different TV stations called "DepEd TV".
If I were to estimate, I can say that these strategies that we implemented were able to help the learners (85%). Nothing is more effective than face-to-face classes. Wherein, we can deliver our lessons effectively and be able to adjust to the variety of learners. We were able to assess them justifiably and accurately.

**Appendix A.4. Mitigations from Consultant 4**

Consultant Name: Gerald John Guillermo
Consultant Expertise: Policy Researcher in the Local Government Unit of Taguig
Title: Local Government Unit Intervention During the COVID-19 Pandemic

| **Mitigation Measures/Decision Nodes** | **Efficacy** | **Impacts** |
|---|---|---|
| Mobile Market project in partnership with the Department of Agriculture, ensuring food accessibility during the COVID-19 pandemic.<br><br>https://metronewscentral.net/taguig/barangay-front/metro-cities/taguig-mobile-market-goes-to-barangay-pinagsama<br><br>https://rmn.ph/mobile-market-ng-taguig-muling-aarangkada-sa-ilang-barangay-nito/ | 85% | *Impacts:*<br>● **Enhanced Food Accessibility:** Residents had improved access to fresh vegetables and other food items, reducing the risk of COVID-19 exposure by avoiding crowded markets.<br>● **Support for Local Farmers:** The initiative provided farmers with a stable platform to sell their goods, ensuring a consistent income during uncertain times.<br>*Mitigation Measures and Efficacy:*<br>● **Open-Air Setup:** Conducting the market in open spaces reduced transmission risks associated with enclosed areas.<br>● **Waste Reduction Strategies:** To address concerns about unsold produce being discarded, the city could implement programs to donate excess food to local shelters or convert it into compost, promoting sustainability.<br>● **Operational Sustainability:** Ensuring the longevity of the mobile market beyond the pandemic requires integrating it into the city's regular services, possibly through public-private partnerships to share operational costs. |
| Robust query-handling system for the Taguig Safe City Task Force Hotline, | 80% | *Impacts:*<br>● **Centralized Information Source:** Residents and visitors had a reliable platform to seek information, reducing misinformation and confusion.<br>● **System Overload Challenges:** The initial setup |
| addressing thousands of inquiries on COVID-19 rules and guidelines | | faced challenges due to high call volumes and limited staffing, leading to delays and the occasional dissemination of incomplete information.<br>*Mitigation Measures and Efficacy:*<br>● **Infrastructure Enhancement:** Expanding the hotline infrastructure by increasing the number of lines and employing additional trained personnel can improve response times and service quality.<br>● **Standardized Training Programs:** Implementing comprehensive training for handlers ensures they are well-equipped to manage a wide range of inquiries professionally, even under pressure.<br>● **Data-Driven Adjustments:** Regular analysis of common queries allows the city to update automated responses and FAQs, streamlining the information dissemination process. |
| Containment and security data, coordinating inter-departmental efforts for localized lockdowns and reporting to the Regional Inter-Agency Task Force (IATF) | 85% | *Impacts:*<br>● **Targeted Containment:** Localized lockdowns minimized disruptions by restricting only specific areas with reported cases, allowing the rest of the city to maintain economic activities.<br>● **Transparency and Compliance:** Regular reporting to the IATF and clear communication with the public fostered trust and adherence to lockdown measures.<br>*Mitigation Measures and Efficacy:*<br>● **Inter-Departmental Coordination:** Establishing a centralized command center facilitated efficient communication and resource allocation among departments during lockdowns.<br>● **Provision of Essential Services:** Ensuring that affected residents received necessary supplies and support during lockdown periods mitigated the adverse effects of movement restrictions.<br>● **Continuous Monitoring:** Utilizing real-time data analytics enabled the city to adjust strategies promptly in response to changing infection patterns. |
| Different avenues for COVID-19 vaccination for different sectors and extensive means to reach out to sectors | 95% | *Impacts:*<br>● **High Vaccination Rates:** The city's proactive approach resulted in one of the highest vaccination rates in the region, contributing to community immunity and a reduction in COVID-19 cases.<br>● **Accessibility and Convenience:** By offering vaccinations in commercial spaces and deploying mobile units, the city ensured that vaccines were accessible to all, including those in hard-to-reach areas.<br>*Mitigation Measures and Efficacy:*<br>● **Public-Private Partnerships:** Collaborating with |

| | | private entities expanded the reach and resources of the vaccination program, enhancing its effectiveness.<br>● **Community Engagement:** Engaging local leaders and organizations in awareness campaigns addressed vaccine hesitancy and encouraged public participation.<br>● **Logistical Planning:** Efficient scheduling systems and adequate staffing ensured smooth operations at vaccination sites, reducing wait times and improving the overall experience. |
|---|---|---|

**Appendix A.5. Mitigations from Consultant 5**

Consultant Name: France Castro
Consultant Expertise: Labor Economist
Title: Labor-Centered Interventions and Efficacy Assessment

| Intervention (Decision Node) | Estimated Efficacy | Expected Impacts |
|---|---|---|
| Emergency Cash-for-Work (CFW) Program at Barangay Level | 90% | Provides immediate income for displaced workers while helping communities with public health needs (e.g., sanitation, repairs to isolation facilities, food distribution).<br><br>Boosts both economic and psychosocial well-being.<br><br>Highly effective in low-income communities with high informality. |
| Barangay-Based Livelihood Reskilling and Cooperative Support | 80% | Empowers unemployed or displaced workers, especially women, with new income streams (e.g., food processing, sewing, community gardening).<br><br>Sustainability improves when tied to cooperatives or LGU procurement.<br><br>Requires minimal start-up capital. |
| Protection of Frontline Education Workers through | 75% | Ensures teachers and para-teachers, especially in public and Lumad schools, |
| Local Safety Audits and PPE Funding | | have sufficient protection to resume or continue duties.<br><br>Helps avoid further educational decline and burnout.<br><br>Reduces COVID-19 transmission in learning environments. |
| Barangay Labor Rights Awareness and Reporting Mechanism | 70% | Establishes safe local spaces for reporting labor violations, including illegal dismissals, wage theft, and lack of benefits.<br><br>Promotes labor protection especially in informal sectors. Stronger when integrated with DOLE and CSO partners. |
| Expanded Barangay Health Workers' (BHW) Livelihood Support + Regularization Push | 70% | Strengthens health systems by formally recognizing and compensating BHWs.<br><br>Increases morale and retention.<br><br>Creates ripple effects on community health, pandemic response, and public trust.<br><br>Requires national-local coordination. |
| Community-Based Daycare and Elder Care Support Services | 80% | Frees up unpaid caregivers (often women) to rejoin the workforce.<br><br>Improves child development and elder welfare.<br><br>Can be facilitated by trained barangay volunteers.<br><br>Key to economic recovery in marginalized households. |
| Reintegration Programs for Returning OFWs in Barangays | 75% | Supports returning migrant workers through livelihood grants, skills matching, and psychosocial services. |
| | | Reduces vulnerability to re-migration under unsafe terms.<br><br>Relies on strong barangay coordination and reintegration planning. |

**Appendix A.6. Mitigations from Consultant 6**

Consultant Name: Bureau of Local Government Development (BLGD), Department of the Interior and Local Government (DILG), Philippines
Consultant Expertise: Local-government policy, disaster-response planning, and development-program monitoring
Title: Local Development Planning Division – COVID-19 Response and Recovery Team

A.6.1 Mandatory Wearing of Face Masks (page 1)

DILG MC No. 2020-071 Mandatory Wearing of Face Masks or Other Protective Equipment in Public Areas, dated 09 April 2020

Said circular translated Inter-Agency Task Force for the Management of Emerging Infectious Diseases (IATF-MEID) Resolution No. 18, dated 01 April 2020, into a directive that also provided for monitoring LGU action/s as it enjoined LGUs in areas placed under Enhanced Community Quarantine to enact local ordinances, executive orders, and other issuances on the mandatory wearing of face masks by all residents to lessen local transmission of COVID-19.

A.6.2 Temporary Shelters for Health Workers (page 1)

DILG MC No. 2020-072 Temporary Shelter/Accommodation for the Safety and Protection Against Discrimination of Health Workers in Provincial/City Hospitals and Other Public Health Facilities Catering to COVID-19 Patients, dated 11 April 2020

Pursuant to Republic Act No. 11469 and the IATF-MEID Resolution No. 19, dated 03 April 2020, the Circular sought to protect public health workers from discrimination by enjoining LGUs to: (i) provide for free/extend assistance in securing temporary shelter/accommodation; (ii) provide incentives to establishments tapped for the purpose; (iii) ensure sanitation management and needs of public health workers in temporary shelters; and (iv) issue appropriate policies/advisories denouncing and providing sanctions and penalties for discriminatory acts against health workers.

A.6.3 Assistance to Relocated Informal-Settler Families (page 2)

DILG MC No. 2020-090 Provision of Assistance to the Relocated Informal Settler Families (ISFs) in Resettlement Communities in Response to the Coronavirus Disease 2019 (COVID-19) Emergency, dated 29 May 2020

Republic Act No. 11469 underscores protection of the vulnerable sectors as Section 3 (b) directs immediate mobilization of assistance in the provision of basic necessities to families and individuals affected by the imposition of Community Quarantine, especially indigents and their families. In line with this provision, the MC directed sending LGUs and receiving LGUs to perform their mutual and shared responsibility of extending financial and/or non financial assistance to ISF relocatees of national government relocation programs in response to the COVID-19 crisis.

A.6.3 Food-Security & Nutrition Information Programme (KUMAIN Series) (page 6)

Participation in the Inter-Agency Task Force on Zero Hunger

Per Executive Order No. 101, dated 10 January 2020, the IATF on Zero Hunger was created to ensure that the government policies, initiatives, and projects on attaining zero hunger shall be coordinated, responsive, and effective. One of the expected outputs of the task force is to come up with a National Food Policy that includes initiatives to end hunger, achieve food security and improve nutrition, and promote sustainable agriculture. The IATF on Zero Hunger is composed of 6 Key Result Areas (KRAs) where in the DILG-BLGD is tasked to lead the National Food Policy KRA 6: Ensure Information, Education, Awareness, and People Participation.

In partnership and coordination with member agencies of KRA 6, a Roadmap on Information, Education, Awareness, and People Participation was drafted and presented through the KUMAIN (Kasapatan at Ugnayan ng Mamayan sa Akmang Pagkain at Nutrition) Series: Stakeholders' Consultation.

Primarily, the objectives of the said KUMAIN Series of consultation with stakeholders are to: 1) inform the stakeholders on the Zero Hunger National Food Policy (NFP); 2) further refine the NFP's KRAs as strategies to address the problem on hunger and nutrition in the country compounded by the effects of the COVID-19 pandemic by gathering feedback and/or possible inputs to the NFPs and its KRAs from the stakeholders; and 3) ensure collaborative efforts and potentially secure commitments thru partnerships among stakeholders in conformity of the NFP's goal. This consultation was participated in by development partners, private sector, academe, religious groups, civil society organizations, non-government organizations, and LGUs.

Series of consultations were conducted on the following dates:

• 26 November 2020 for development partners

• 27 November 2020 for private sector

• 01 December 2020 for academe

• 03 December 2020 for religious groups, CSOs, NGOs

• 04 December 2020 for LGUs